\documentclass{epl}
\usepackage{amssymb}

\title{Hard colloidal rods near a soft wall: wetting, drying, and symmetry breaking}
\shorttitle{Hard colloidal rods near a soft wall: ...}
\author{Kostya Shundyak\inst{1} \and Ren\'{e} van Roij\inst{2}}
\institute{
  \inst{1} Instituut-Lorentz for Theoretical Physics, Leiden University -
  Niels Bohrweg 2, Leiden, NL-2333 CA, The Netherlands.\\
  \inst{2} Institute for Theoretical Physics, Utrecht University -
  Leuvenlaan 4, 3584 CE Utrecht, The Netherlands.
}

\begin{document}

\maketitle

\begin{abstract}
Within an Onsager-like density functional theory we explore the
thermodynamic and structural properties of an isotropic and
nematic fluid of hard needle-like colloids in contact with a hard
substrate coated with a soft short-ranged attractive or repulsive
layer.  As a function of the range and the strength of the soft
interactions we find wetting and drying transitions, a pre-drying
line, and a symmetry-breaking transition from uniaxial to biaxial
in the wetting and drying film.
\end{abstract}

Whereas bulk liquid crystal phases of suspensions of colloidal
hard rods have been essentially understood due to Onsager's work
in the 1940's \cite{O,VLRPP92} and simulations \cite{BFJCP97} and
density functional theories \cite{STPRL88,PHPRL88,MSPRA89} in the late 1980's,
their surface and interfacial properties are subject of ongoing
study, not only experimentally but also theoretically and
numerically. Good progress was made during the past decade in the
theoretical study of planar free isotropic-nematic (IN)
interfaces, e.g. it is known by now that the nematic director in
the thermodynamically stable state is parallel to the interfacial
plane \cite{CN92,AJCP00,SRJPCM01}, that complete wetting of the IN
interface by another nematic phase occurs near the triple point in
binary rod mixtures \cite{SRPRL02}, and that the one-particle
distributions in the IN interface are only weakly biaxial
\cite{CPRE93,SRPRE03,WTSRS}. The effect of external substrates on
suspensions of hard rods has also been studied. For a planar hard
wall, for instance, evidence of complete wetting of the
wall-isotropic (WI) interface by an intervening nematic film was
provided by theory \cite{PHPRA88} and simulations \cite{DREPRE01}.
Other studies were concerned with properties of a hard-rod fluid
in contact with a ``penetrable'' wall, which restricts only the
translational degrees of freedom of the rods \cite{AMP99}. It was
shown that such a wall favors homeotropic anchoring of the nematic
director, and that the WI interface exhibits complete wetting by
the homeotropically aligned nematic phase upon approach of IN
coexistence \cite{AJCP00,HMVPRE03}. The common feature of all
these studies is that the chosen wall potential does not allow to
control the degree of surface nematic order. This is in contrast
with Landau-de Gennes theory, which predict rich surface phase
diagrams for liquid crystals \cite{SPRL76,SPPRL85,JRPP91}. A
drawback of this formalism is, however, that the effects of
particular surface-particle interactions are hidden in expansion
coefficients that are not always easily traced back to the
microscopic details \cite{SSPRA86,JRPP91}. Therefore, it is interesting
and relevant to construct surface phase diagrams, for a ``simple''
fluid of monodisperse hard rods, in terms of parameters of the
surface potential.

Calculating surface phase diagrams of colloidal liquid crystals
from a microscopic theory is technically involved and
computationally expensive, as density profiles
$\rho(z,\theta,\varphi)$ are to be determined for a given wall
potential $V(z,\theta,\phi)$, with $z$ the distance from the wall
and $\theta$ and $\varphi$ the polar and azimuthal angle of the
orientation of the rod. Building on earlier studies
\cite{TdGMP84,PRVMRPRL99}, only recently a study appeared of the
surface properties of a hard-spherocylinder fluid in contact with
a model substrate, composed of a ``penetrable'' wall and a
short-ranged repulsive or attractive tail \cite{HMVPRE03}. Several
wetting transitions and a transition from homeotropic to planar
anchoring were observed at different strengths of the wall
potential. This can be understood qualitatively, since the
``penetrable'' wall with the attractive tail favors homeotropic
alignment of the nematic director, and for the strong repulsive
tail it resembles the hard wall such that the stable director
changes to planar. However, the complexity of the model prohibits
to explore the full parameter space.

In order to be able to map a complete surface phase diagram we
sacrifice the full rotational symmetry of the rods and consider
the simpler Zwanzig model, with a restricted number of allowed
orientations \cite{ZJCP63}. This model exhibits a strong
first-order IN transition in the bulk \cite{ZJCP63}, and the
orientation of the nematic director parallel to the IN interface
was found to be thermodynamically favorable \cite{RDEEL00},
similar to continuous rods. In contact with a hard wall the WI
interface is completely wet by the N phase upon approach of the IN
coexistence, whereas the WN interface remains partially wet
\cite{RDEEL00}, i.e. in such a geometry the model also shows
behavior similar to continuous rods. Although some limitations of
the Zwanzig model are known \cite{SJCP72,SRPRE04}, we adopt it for
the present study for reasons of its numerical simplicity. We
explore the complete surface phase behavior of a Zwanzig hard-rod
fluid in contact with a substrate which consists of a hard wall
and a short-ranged attractive or repulsive tail. In contrast with
the ``penetrable'' wall it puts the infinite potential barrier at
the origin regardless the orientation of the rod.

We consider an inhomogeneous fluid of rectangular hard rods of
length $L$ and diameter $D$ ($L\gg D$) in a macroscopic volume $V$
at inverse temperature $\beta$ and chemical potential $\mu$. The
rod orientations are restricted to the three mutually
perpendicular directions $\hat{n}_i$ representing $\hat{x}$,
$\hat{y}$, $\hat{z}$ for $i=1,2,3$, respectively. The position
${\bf r}$ of the center of mass of a rod is continuous. The grand
potential functional of the fluid in an external potential
$V_i({\bf r})$ with the one-particle distribution functions
$\rho_i ({\bf r})$ can be written, within the second virial
approximation, as \cite{RDEEL00}
\begin{eqnarray}\label{omega}
\beta\Omega[\rho]=\sum_{i=1}^3 \int d{\bf r}
\rho_i({\bf r}) \Big(\ln [\rho_i ({\bf r}) \nu]
-1-\beta\mu + \beta V_i(\mathbf{r}) \Big)
-\frac{1}{2}\sum_{i,j=1}^3 \int d{\bf r}d{\bf r}' f_{ij}({\bf
r};{\bf r}') \rho_i({\bf r})\rho_j({\bf r}'),
\end{eqnarray}
with $\nu=L^2D$, and $f_{ij}({\bf r};{\bf r}')=-1, (0)$ the Mayer
function of two (non-)overlapping hard rods with orientations
$\hat{n}_i$ and $\hat{n}_j$ and center-of-mass coordinates ${\bf
r}$ and ${\bf r}'$.
Qualitatively, the second virial functional predicts the bulk IN
transition and the adsorption and wetting properties near a planar
hard wall in agreement with simulations of freely rotating rods
\cite{RDEEL00}.

The external potential $V_i({\bm r})$ consists of a single planar
hard wall in the plane $z=0$ (normal $\hat{z}$) and a generic soft
"tail" $\propto\exp(-\kappa z)$ acting on each rod segment, and is
written as
\begin{eqnarray}
\label{softwallpot} \beta V_i(\mathbf{r})&=& \left\{
\begin{array}{ll}
\beta A \exp [-\kappa z]  & z>0, i=1,2;\\
\beta AF \exp [-\kappa z] &z>\frac{L}{2}, i=3;\\
\infty  &\mathrm{otherwise},
\end{array}
\right.
\end{eqnarray}
where the form factor $F=\sinh(\kappa L/2)/ (\kappa L/2)>1$ arises
from integration over the contour of the rod. For $A>0$ such a
potential could , under certain circumstances, be caused by a
polymer coating, with $A$ a measure for the planar density and
$\kappa^{-1}$ for the radius of gyration \cite{deGennes}. In the
present study we avoid discussions on the microscopic origin of
$V_i({\bf r})$, and explore the generic properties of the surface
phase diagram as a function of the ``contact'' potential $A$ and
the decay constant $\kappa$. Our primary interest is in the
parameter regime $\beta A \leq 10$, since larger values of $\beta
A$ correspond essentially to a shift of the hard wall along the
$z$-axis.

\begin{figure} \centering
\includegraphics*[width=6.5cm]{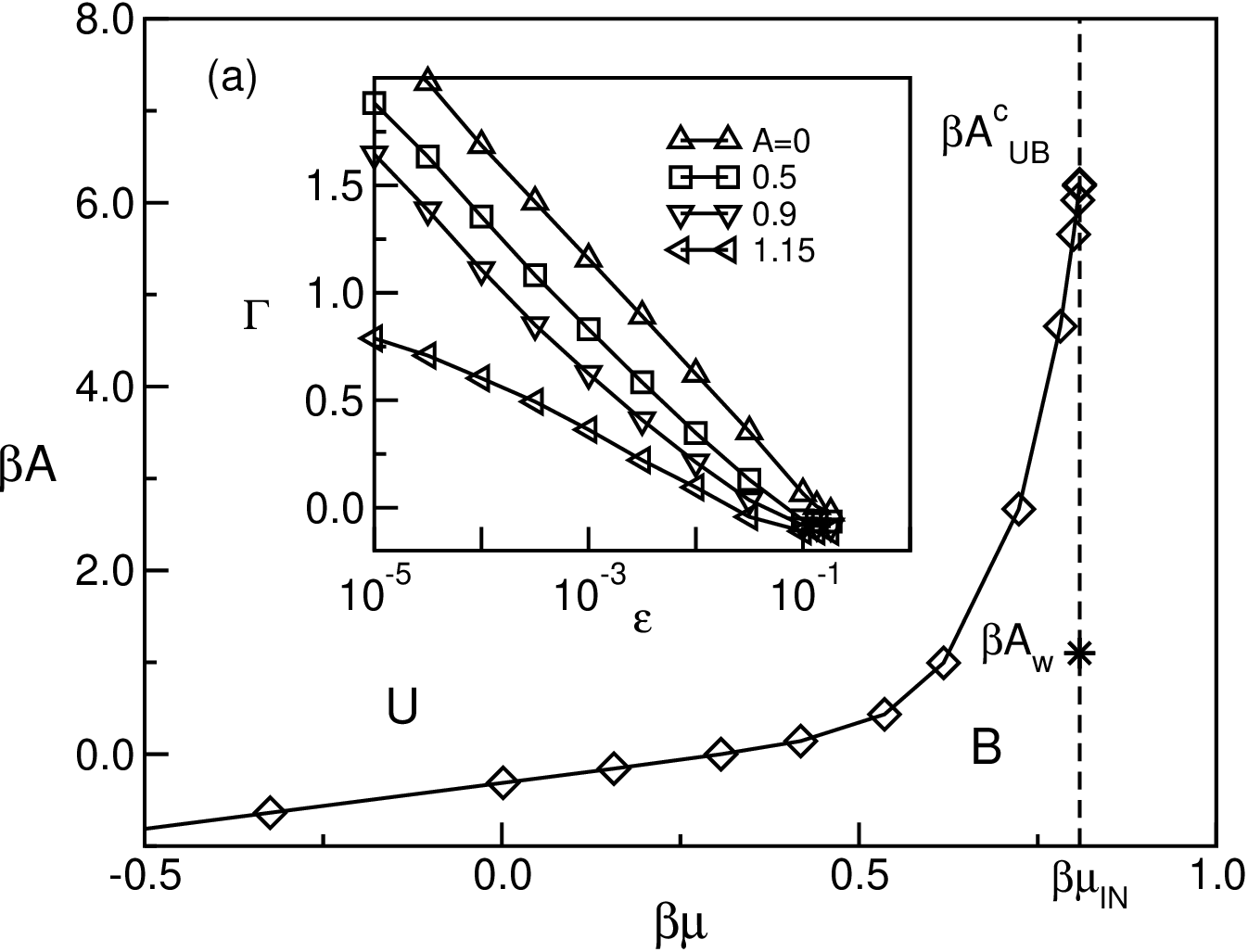}\,\,\,
\includegraphics*[width=6.5cm]{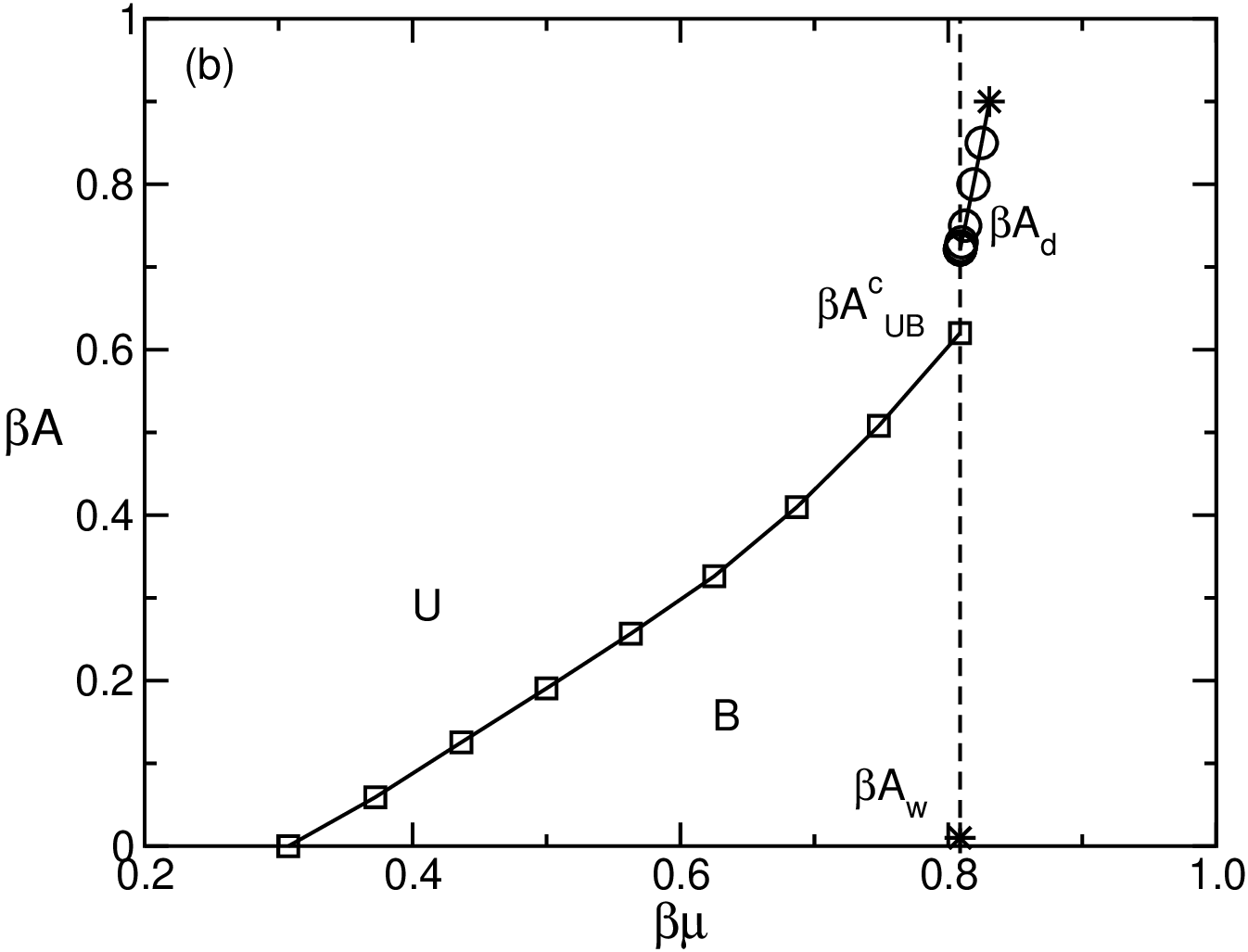}\\
\vspace{0.2cm}
\includegraphics*[width=6.5cm]{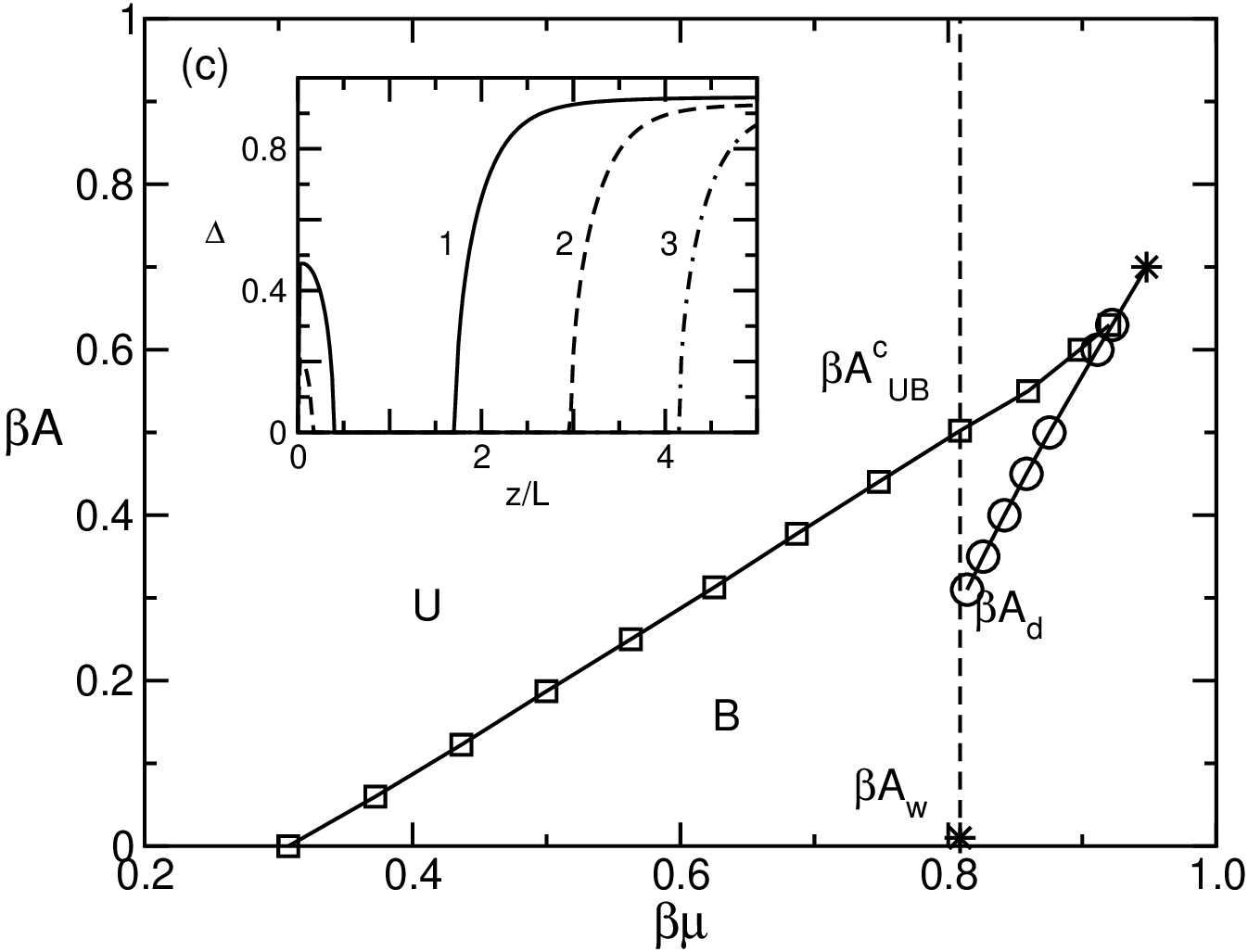}\,\,\,
\includegraphics*[width=6.5cm]{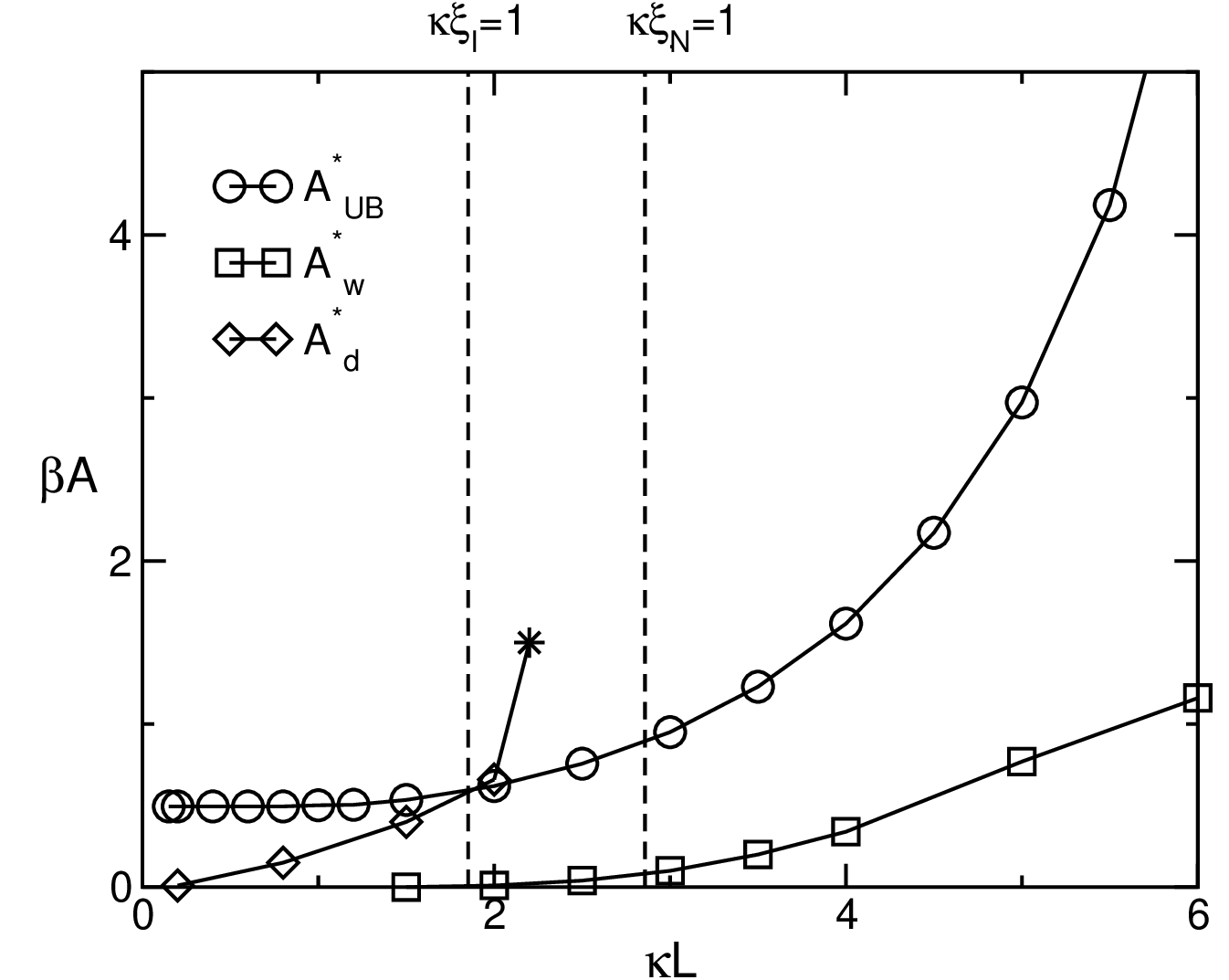}
\caption{ \label{phasediag} \small (a,b,c) Surface phase diagrams
of the Zwanzig hard-rod fluid in contact with a soft wall in terms
of the contact potential $\beta A$ and chemical potential $\beta
\mu$ for different range of the wall potential $\kappa L=6$ (a),
$\kappa L=2$ (b), and $\kappa L=1$ (c). In all three cases we
distinguish a uniaxial ($U$) and biaxial ($B$) phase for $\mu
<\mu_{IN}$, and a wetting transition ($\ast$) at $\mu =\mu_{IN}$
that separates complete wetting of the wall by a nematic film at
$A<A_w$ from partial wetting $A>A_w$. In (b) and (c) we also see a
predrying line at $\mu>\mu_{IN}$, and a drying transition at
$\beta \mu = \beta \mu_{IN}$, that separates complete drying by an
isotropic film at $A>A_d$ from partial drying $A<A_d$. The inset
in (a) illustrates the logarithmic divergence of the adsorption
$\Gamma$ with undersaturation $\epsilon=1-c_b/c_I$ for $A<A_w$,
and that in (c) shows several biaxiality profiles $\Delta(z)$ for
$A=0.55>A^c_{UB}$ for several values of the bulk chemical
potential $\mu_{1,2,3}>\mu_{IN}$. (d) Generic surface phase diagram for
a fluid of Zwanzig hard rods {\it at $IN$ coexistence} in contact
with the ``soft'' wall. Solid lines correspond to the wetting
($\square$), drying ($\diamond$), and the UB ($\circ$)
transitions; the end point on the line of drying transitions is
indicated by ($\ast$). The dashed lines indicate
$\kappa\xi_{I,N}=1$, with the bulk correlation lengths
$\xi_I/L=0.54$ and $\xi_N/L=0.35$ of the bulk I and N  phase,
respectively.}
\end{figure}

Throughout we assume translational invariance in the $xy$ plane,
such that the dimensionless densities $c_i(z)=\rho_i({\bf
r}){\nu}$ only depend on the distance from the substrate $z$. The
Euler-Lagrange equations $\delta\Omega[\{\rho \}]/
\delta\rho_i({\bf r})=0$ can then be written as \cite{RDEEL00}
\begin{eqnarray}
\label{el} \beta \mu = \ln c_1(z) +\beta V_1(z) +2
c_2(z)+2\bar{c}_3(z), \nonumber\\
\beta \mu = \ln c_2(z) +\beta V_2(z) +2 c_1(z)+2\bar{c}_3(z), \\
\beta \mu = \ln c_3(z) +\beta V_3(z) +2 \bar{c}_1(z) +2
\bar{c}_2(z), \nonumber
\end{eqnarray}
with the averaged densities $\bar{c}_i(z)=(1/L) \int_{-L/2}^{L/2}
dz' c_i(z+z')$. We solve Eqs. (\ref{el}) iteratively for a given
$A$ and $\kappa$ at a fixed chemical potential $\mu$. In all
numerical calculations we use an equidistant $z$-grid of $40$
points per $L$. Convergence is assumed when the relative
difference between the results of the subsequent iterations is
smaller then $10^{-10}$ for all values of $z$ in the grid. Such
accuracy is required to avoid dependencies of the results of the
calculations on the initial guesses. Additional checks, performed
with 80 points per $L$, yield virtually identical results. Once
determined, the equilibrium profiles can be inserted into the
functional to obtain the equilibrium value of the grand potential
\begin{eqnarray}
\label{potmin} \beta\Omega=\frac{{\cal A}}{2{\nu}} \sum_i\int dz
c_i(z)\Big(\ln[c_i(z)]-2-\beta\mu + \beta V_i(z) \Big).
\end{eqnarray}
Note that $\Omega=-pV$ for a bulk system in a volume $V$, with
$p=p(\mu,T)$ the pressure. In the presence of a planar surface or
interface of area ${\cal A}$ we have $\Omega=-pV+\gamma {\cal A}$
with $\gamma=\gamma(\mu,T)$ the surface or interface tension. The
results of the calculations can be conveniently represented in
terms of the total density $c(z)=\sum_{i=1}^3 c_i(z)$, the nematic
order parameter $s(z)=(c_3(z)-\frac{1}{2}(c_1(z)+c_2(z)))/c(z)$,
and the biaxiality $\Delta(z)=(c_1(z)-c_2(z))/c(z)$. The
(dimensionless) adsorption near a planar substrate is defined as
$\Gamma(\mu) =\int_0^\infty dz (c(z)-c_b)/L$, where $c_b=c_b(\mu)$
is the bulk density far from the substrate. In
Fig.~\ref{phasediag}(a,b,c) we show $\mu-A$ surface phase diagrams
for several ranges of the wall potential $\kappa L=6$ (a), $\kappa
L=2$ (b), and $\kappa L=1$ (c). The phase diagrams are very rich,
therefore we discuss separately their parts which correspond to
the low ($\mu\leq \mu_{IN}$) and high ($\mu>\mu_{IN}$) values of
the bulk chemical potential.

For $\mu\leq \mu_{IN}$ all three phase diagrams of
Fig.~\ref{phasediag}(a,b,c) exhibit a uniaxial-to-biaxial (UB)
surface transition at $\mu=\mu_{UB}(A)<\mu_{IN}$ for small enough
$A<A_{UB}^c$. This implies that only sufficiently repulsive walls
do {\em not} show the UB transition, such that $\Delta(z)\equiv 0$
even upon approach of $IN$ coexistence. The critical value
satisfies $\beta A_{UB}^c>0.5$ in all cases, and increases with
$\kappa L$ (Fig.~\ref{phasediag}(d)). The $UB$ transition is found
to be continuous for all investigated combinations of $\kappa L$
and $\beta A$. This is consistent with the findings of Ref.
\cite{RDEEL00}, where the case $\beta A=0$ (hard-wall) was
considered.

\begin{figure}[t]\centering
\includegraphics*[width=6.9cm]{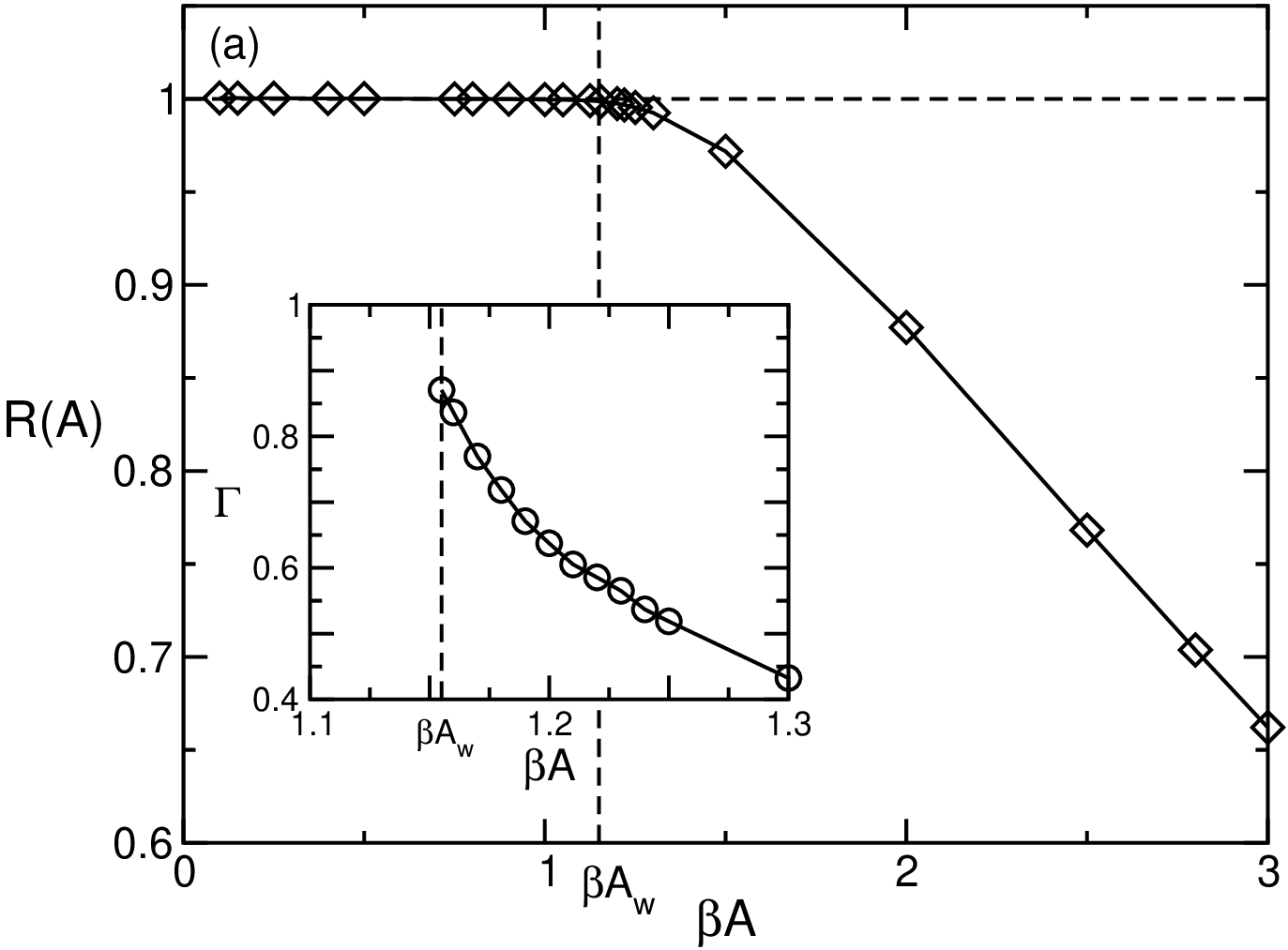}\,\,\,
\includegraphics*[width=6.5cm]{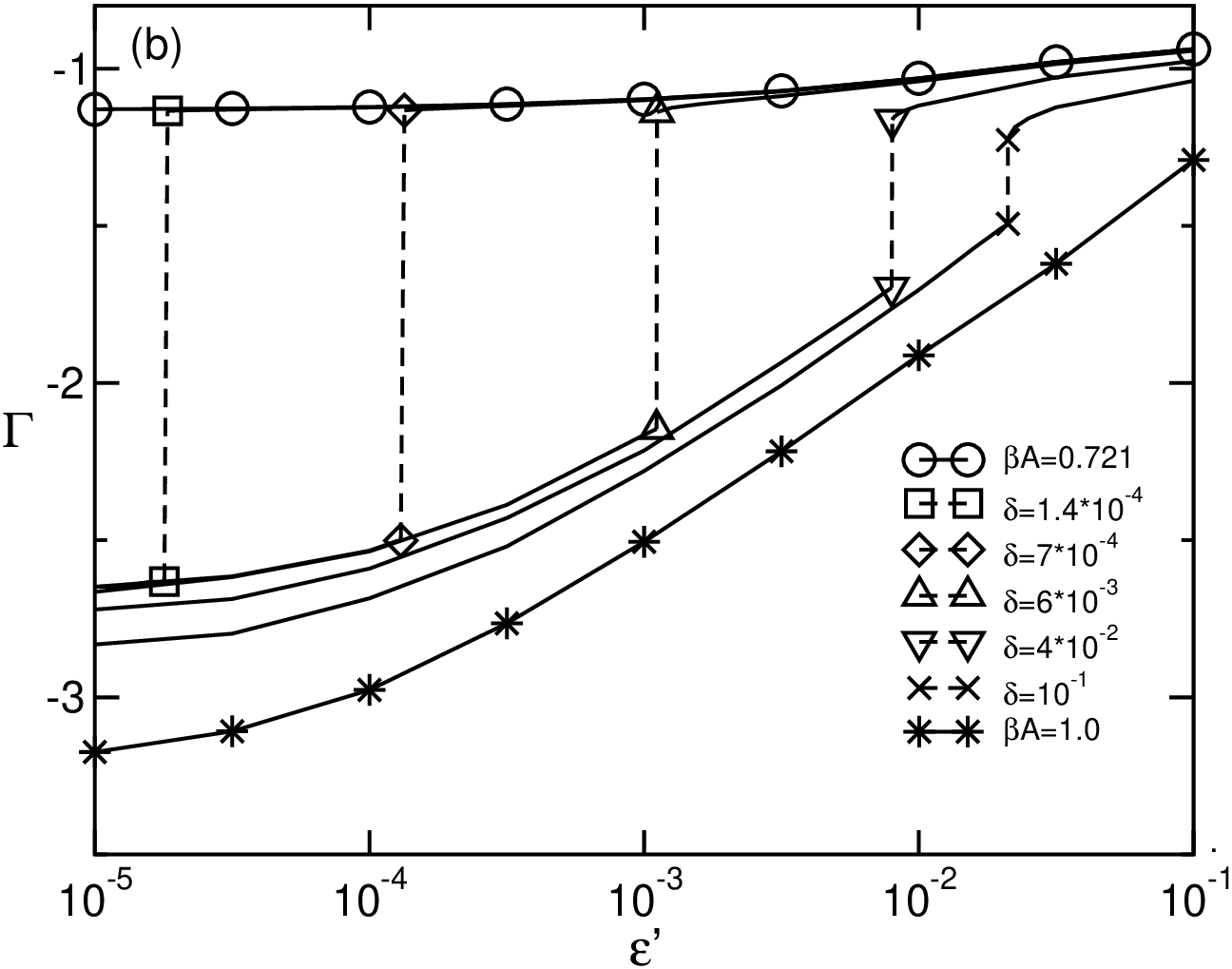}
\caption{\label{RGamma} \small (a) The relative surface tension
difference $R(A)$ at the $WI$ and $WN$ interfaces for a wall
potential with a decay given by $\kappa L=6$. The vertical dashed
line $\beta A_w=1.15$ indicates the repulsion amplitude at which
the wetting transition takes place. The inset shows the excess
adsorption $\Gamma (A)$ at coexistence. (b) Excess desorption
$\Gamma(\epsilon')$ at the interface between the $N$ phase and the
soft wall ($\kappa L=2$) as a function of the oversaturation
$\epsilon'=c_b/c_N-1$ for several different values of $\beta
A=0.721(1+\delta)$. The vertical dashed lines connect coexisting
points at the predrying line.}
\end{figure}

For all studied $\kappa L$ we only recover the hard-wall complete
wetting phenomenon \cite{RDEEL00}) for sufficiently weak
repulsions $A<A_w$, whereas strong repulsions $A>A_w$ give rise to
partial wetting. This is illustrated in the inset of Fig.1(a),
where the adsorption $\Gamma$ diverges logarithmically with the
undersaturation $\epsilon=1-c_b/c_I\rightarrow 0$ for $A<A_w\simeq
1.1$, and remains finite for $A>A_w$. Here $c_I$ is the coexisting
isotropic bulk density.  This implies that a wetting transition
takes place at $A=A_w$. Interestingly, the numerical value
$A_w\simeq 10^{-2}$ is seen to be very small for longer-ranged
potentials, $\kappa L\leq 2$, i.e. extremely weak but sufficiently
long-ranged repulsions strongly reduce the complete-wetting
regime. Shorter-ranged substrate potentials exhibit a larger
complete wetting regime, e.g. $A_w=\simeq 1.1$ when $\kappa L=6$
(see also Fig.\ref{phasediag}(d)). The existence of a wetting
transition at $A=A_w$ follows also from comparisons of the
corresponding interfacial energies. For convenience, we define the
relative surface tension difference
\begin{eqnarray}
\label{Rfunc} R(A) = \frac{\gamma_{WI}(A) - \gamma_{WN}(A)}{
\gamma_{IN}},
\end{eqnarray}
with $\gamma_{WI}(A)$ and $\gamma_{WN}(A)$ the tensions of the
$WI$ and $WN$ interface, respectively, both at $\mu=\mu_{IN}$, and
$\beta \gamma_{IN}LD=(2.8027\pm 0.0001)\times 10^{-2}$ the surface
tension of the free planar $IN$ interface \cite{RDEEL00}. In
Fig.~\ref{RGamma}(a) we show $R(A)$ for $\kappa L=6$. It can be
seen that $R(A)\equiv 1$ (with relative accuracy of $10^{-3}$) if
$\beta A<\beta A_w=1.15$, and $R(A)<1$ if $A>A_w$. This is a clear
thermodynamic evidence for a transition from partial to complete
wetting at $A=A_w$.

The study of the order of the wetting transition represents
significant numerical difficulties. Our results suggest that it is
likely a first-order wetting transition, because  (i)
$R(A)$ appears to approach $R(A_w)=1$ with a finite slope when
$A\downarrow A_w$, (ii) metastable nematic films of finite
thickness can easily be generated at $\mu=\mu_{IN}$ for $A<A_w$,
and (iii) the adsorption $\Gamma(\mu_{IN})$ at saturation appears
to show a discontinuous jump from a finite value for $A>A_w$ to an
infinite value for $A<A_w$ (see insets Fig.~\ref{phasediag}(a) and 
\ref{RGamma}(a)). One expects, for reasons of continuity, that a first-order wetting
transition is accompanied by a first-order prewetting transition,
from a finite thin to a finite thick film off coexistence.
Unfortunately, we have not been able to find this prewetting
transition, despite considerable efforts. We speculate that its
critical point is too close to the wetting transition to be
detected.

We now consider the $\mu>\mu_{IN}$ part of the phase diagrams,
where the $N$ bulk phase has director $\hat{x}$, i.e. planar. We
have checked that a homeotropic orientation of the director is
metastable in all the cases studied. Intuitively, one expects a
low-density isotropic film close to a highly repulsive substrate
($\beta A\gg1$), which may but need not become of macroscopic
thickness when $\mu\downarrow\mu_{IN}$. For $\beta A=0$ it is
known that $\Gamma<0$ but finite \cite{RDEEL00}, a situation that
we refer to as partial drying. Our numerical results for $\kappa
L=6$ show that $\Gamma$ remains finite when
$\mu\downarrow\mu_{IN}$ for any $\beta A<10$. For $\beta A \geq
10$ the density of rods close to the hard wall ($z=0$) is so small
that its position can be shifted to some $z=z_0>0$ without
affecting the profiles $c_i(z)$ significantly, while the effective
amplitude is reduced from $A$ to $A\exp(-\kappa z_0)\leq 10$. We
can conclude that the soft potential with $\kappa L=6$ is so
short-ranged that the partial drying is obtained for any value
of $\beta A$. Fig.1(a) is therefore featureless for $\mu\geq \mu_{IN}$.

The situation is more interesting for the longer-ranged cases
$\kappa L=2,1$, where we do find a drying transition at $\beta
A_d=0.72, 0.31$ in Fig.~\ref{phasediag}(b), (c), respectively.
When $\mu \downarrow \mu_{IN}$ partial drying is observed for
$A<A_d$, and complete drying by an isotropic film for $A>A_d$. In
Fig.~\ref{RGamma}(b) we show  the excess adsorption
$\Gamma(\epsilon')$ as a function of oversaturation
$\epsilon'=c_b/c_{N}-1$ for various amplitudes of the wall
potential with $\kappa L=2$. At low values of the contact
potential ($A \leq A_d$) the desorption remains almost constant
upon approach of the $IN$ coexistence. Calculations show that
$R(A)>-1$ in this regime, corresponding to partial drying. The
behavior changes qualitatively for $0.72<\beta A<0.95$, where
$\Gamma (\epsilon')$ exhibits a discontinuity at some
$\epsilon'>0$, which we associate with predrying. For $\beta
A\simeq 0.95$ the jump of the desorption takes place at
$\epsilon'=0.05$, and its magnitude is (vanishingly) small, i.e.
this is the critical predrying point indicated by ($\ast$) in Fig.
\ref{phasediag}(b) and (c). The jump of the desorption at the
predrying line increases upon decreasing $\beta A$ towards $\beta
A_d$, and its location shifts to smaller $\epsilon'$, until it
diverges at the drying transition at $\beta A=\beta A_d$ at
$\epsilon' \rightarrow 0$. For $\beta A \geq 1.0$, i.e. above the
critical predrying amplitude, the desorption grows continuously
upon approach of coexistence. Studies of the surface tensions at
coexistence show that $R(A)=-1.000\pm 0.001$ for all $A> A_d$,
consistent with a complete drying of the $WN$ interface by an
intervening isotropic film. However, the profiles (not shown) do
not exhibit a clear bulk film even at $\epsilon'=10^{-5}$, which
implies that the asymptotic limit $\epsilon' \rightarrow 0$ has
not been reached yet such that the regime of logarithmic growth of
$-\Gamma$ with $\epsilon'$ cannot properly be identified in Fig.
\ref{RGamma}(b). Interestingly, for $\kappa L \leq 1.5$ we find
that $A_d<A_{UB}^c$, such that for $A>A_d$ the (thick) isotropic
film that grows in between the wall and the nematic bulk phase can
exhibit a UB transition. In other words, the UB transition at
$\mu<\mu_{IN}$ extends continuosly to the thick isotropic films at
$\mu>\mu_{IN}$, as illustrated for $\kappa L=1$ in Fig.
\ref{phasediag}(c). Here the line of $UB$ transitions terminates
at $\beta A^{*}_{UB}=0.63$, where it crosses the predrying line.
We note, finally, that no new physics is to be expected for larger
$A$ in Fig.\ref{phasediag}(b) and (c), since the regime of
approximate scaling of the density profiles (associated with a
shift of the hard wall and a renormalization of the contact
potential) at large $A$ is observed already for $\beta A \geq
1.5$.

All our findings are summarized in Fig. \ref{phasediag}(d), which
represents the generic surface phase diagram in the $\beta A -
\kappa L$ coordinates for a fluid of Zwanzig hard rods at $IN$
coexistence ($\mu=\mu_{IN}$) in contact with the ``soft'' wall. We
distinguish lines of the wetting ($\square$) and drying
($\diamond$) transitions. The $UB$ transition ($\circ$) at the
$WI$ interface occurs for all $\kappa L$. It also takes place at
the $WN$ interface in the thick isotropic wetting film, i.e. below
the intersection of the lines of the drying and the $UB$
transitions. Note that the growth of the wetting film is
determined by the correlation length $\xi_I$ ($\xi_N$)
of the coexisting bulk isotropic and
nematic phase, and no longer by $\kappa^{-1}$,
when $\kappa ^{-1}<\xi_I$ ($\xi_N$).
However, this crossover does not
introduce any additional structure to the surface phase diagram.
For large $\kappa L$ the $UB$ line shows an exponential divergence
$\beta A \sim \exp [\kappa L]$, associated with the
renormalization of the contact potential in the scaling regime
$\beta A =\beta A_0 \exp[\kappa L z_0/L]$. The line of the wetting
transitions is seen to be linear with $\kappa L>4$.  This is
associated with the fact that for the steep potentials density
variations within the decay length of the potential are rather
small, and can be neglected. Therefore, its contribution to the
grand potential is proportional to $(\beta A/L) c(0) \int dz \exp
[-\kappa z]=(\beta A/\kappa L) c(0)$. Since the wetting transition
for large $\kappa L$ occurs at similar values of the surface free
energy, $(\beta A_w/\kappa L) c(0)=\mathrm{const}$, and hence
$\beta A_w \sim \kappa L$.

We conclude that the surface phase behaviour of hard rods with
soft walls is extremely rich, at least on the basis of the present
analysis within the Zwanzig model. We find that complete wetting
of the WI interface by a nematic film, as found for a planar hard
wall, is very sensitive to weak but long-ranged wall repulsions,
and that strongly repulsive walls show complete drying of the WN
interface by an intervening isotropic film if the range of the
potential is long enough. It is of interest to extend the present
study to other related systems, e.g. freely rotating rods or
platelets. Another related interesting question concerns the
required nature of the surface potential that can lead to a stable
anchoring transition in a colloidal-rod fluid \cite{PRVMRPRL99}.
Work along these lines is in progress, and hopefully stimulates
the study of colloidal analogues of liquid crystals near
interfaces.

\acknowledgments
It is a pleasure to thank M. Dijkstra, R. Evans and C. Likos for
stimulating discussions. This work is part of the research program
of the ``Stichting voor Fundamenteel Onderzoek der Materie
(FOM)'', which is financially supported by the ``Nederlandse
organisatie voor Wetenschappelijk Onderzoek (NWO)''.

\end{document}